\documentclass[]{spie}
\usepackage[]{graphicx}

\title{Imaging Gold Nanoparticles in Living Cells Environments using Heterodyne Digital Holographic Microscopy}

\author{Nilanthi Warnasooriya\supit{1,$\dag$}, Fadwa Joud\supit{$\ast$2,4,$\dag$}, Philippe Bun\supit{3}, Gilles Tessier\supit{2},\\
Maite Coppey-Moisan\supit{3}, Pierre Desbiolles\supit{2}, Michael Atlan\supit{1}, Marie Abboud\supit{4}, Michel Gross\supit{2}
\skiplinehalf
\supit{\dag}These authors have equally contributed in this paper\\
\supit{1}~\'Ecole Sup\'erieure de Physique et de Chimie Industrielles de la Ville de Paris, Institut Langevin, UMR 7587
CNRS, 10 rue Vauquelin, 75231 Paris Cedex 05; France\\
\supit{2}~Laboratoire Kastler Brossel - \'Ecole Normale Sup\'erieure , UMR 8552 , UPMC, CNRS\\
24 rue Lhomond , 75231 Paris Cedex 05; France\\
\supit{3};~D\'epartement de Biologie Cellulaire - Institut Jacques
Monod, UMR 7592, CNRS,\\Univ. Paris 6 and 7,  2 Place Jussieu,
75251 Paris Cedex 05;  France \\
\supit{4}~Facult\'e des Sciences - Universit\'e Saint Joseph , B.P :11-514 Riad El Solh,\\Beirut 1107 2050; Lebanon
}

\authorinfo{\supit{$\ast$}~fadwa.joud@lkb.ens.fr}

\begin{document}
  \maketitle

\begin{abstract}
This paper describes an imaging microscopic technique based on heterodyne digital holography where   subwavelength-sized
gold colloids can be imaged in cell environment. Surface cellular receptors of 3T3 mouse fibroblasts are labeled with
40 nm gold nanoparticles, and the biological specimen is imaged in a total internal reflection configuration with
holographic microscopy. Due to a higher scattering efficiency of the gold nanoparticles versus that of cellular structures, accurate localization of a gold marker is obtained within a 3D mapping of the entire sample's scattered field, with a lateral precision of 5 nm and 100 nm in the x,y and in the z directions respectively, demonstrating the ability of holographic microscopy to locate nanoparticles in living cells environments.

\end{abstract}


\section{INTRODUCTION}
\label{sec:intro}  

With recent developments in the fields of nanotechnology  and modern optical microscopy, the use of nanometric
particles as biomarkers in biological specimens has been rapidly increased. Colloidal gold nanoparticles have gained
popularity over other biomarkers for several reasons.
First, gold nanoparticles provide high scattering efficiencies \cite{jain2006}. Second, unlike fluorescent markers,
they are immune to photo bleaching, and they are potentially non-cytotoxic, which provides a great advantage over
semiconductor nanoparticles \cite{west06}. Because of these  properties, the potential for the use of gold
nano particles as biomarkers for live cell imaging using photothermal tracking \cite{lasne2006, cognet2002, boyer2003} and light scattering spectroscopy \cite{raschke2003} has
been demonstrated. Detection of live oral cancer cells using gold nanoparticles have also been obtained \cite{elsayed2005spr}. In this paper, we show the potential of using Digital Holographic Microscopy (DHM) as a powerful
tool to detect and localize, in three dimensions, gold nanoparticles attached to living cells.

In digital holography  a charged coupled device detector (CCD) records the interference diagram, called hologram, of
two waves: the reference wave and the object wave \cite{Schnars_2002}. The interference pattern is then numerically
reconstructed to obtain information about the object wave field. A single hologram can be used to reconstruct the wave
field at virtually any plane, thus allowing one to obtain both amplitude and phase information of the whole object.
Over the past few decades, different configurations and techniques have been introduced in  digital holography,
resulting in precise phase shift \cite{atlan2007aps}, high resolution \cite{leclerc2001,di2008, carl2004},
and high sensitivity measurements \cite{gross2007dhu}. Some of these techniques involve improvement of the experimental setup
\cite{Leclerc2000} and data acquisition methods, while others involve additional numerical treatments of the acquired
holograms and modification of the numerical reconstruction procedure \cite{cuche2000sfz}. In this vision, Leith and
Upatnieks \cite{Leith65} suggested an off-axis configuration holography, in which the combination of the two beams is
done after angularly tilting the propagation direction of one of the beams with respect to the other. Phase-shifting
interferometry adapted to digital holography \cite{yamaguchi1997psd} has been widely used, and our group has proven the
potential efficiency of combining off-axis and phase shifting with a spatial filtering technique for reducing noise
\cite{gross2007dhu}, and eliminating images aliases and overlapping \cite{gross2008naa}. The progress in various
digital holographic techniques has dramatically increased its applications to a large variety of fields, and especially
in the fields of cell biology and biomedical microscopy \cite{xu2001,mann2005,mann2006,charrire2006}.

Holography has now proved its ability to localize scattering nanoparticles in 3D, as shown by Atlan et al. in a recent
paper \cite{atlan2008}, either for fixed particles spin coated on a glass substrate or in free motion within a water
suspension. More recently, Absil et al.\cite{absil2009photothermal} have shown that heterodyne holography also allows the
photothermal detection of small nanoparticles.
However, in order to apply these techniques to biological specimens, important issues must be considered. In biological
samples, the particle holographic signal is superimposed with the light scattered by cell refractive index
fluctuations, which yield a  speckle field. While in our case this speckle is a parasitic signal, in many other
situations, like in Dark Field microscopy, or in Differential Interference Contrast (DIC) microscopy
\cite{goldberg1986saf}, this speckle is the main source of contrast used to image the cell. In this paper, we study
the possibility of 3D holographic imaging in a biological context and we image for the first time 40 nm gold
nanoparticles attached to living cells using DHM. Since the cell scattered speckle field cannot be avoided, it is important for future
cell labeling applications, to scale the particle signal with respect to the scattered speckle. We show here that the
amplitude of the 40 nm gold particle signal we get is much larger than the cell scattered field.

\section{Sample preparation}\label{section_Sample preparation}

\begin{figure}
\centering\includegraphics[width=8cm]{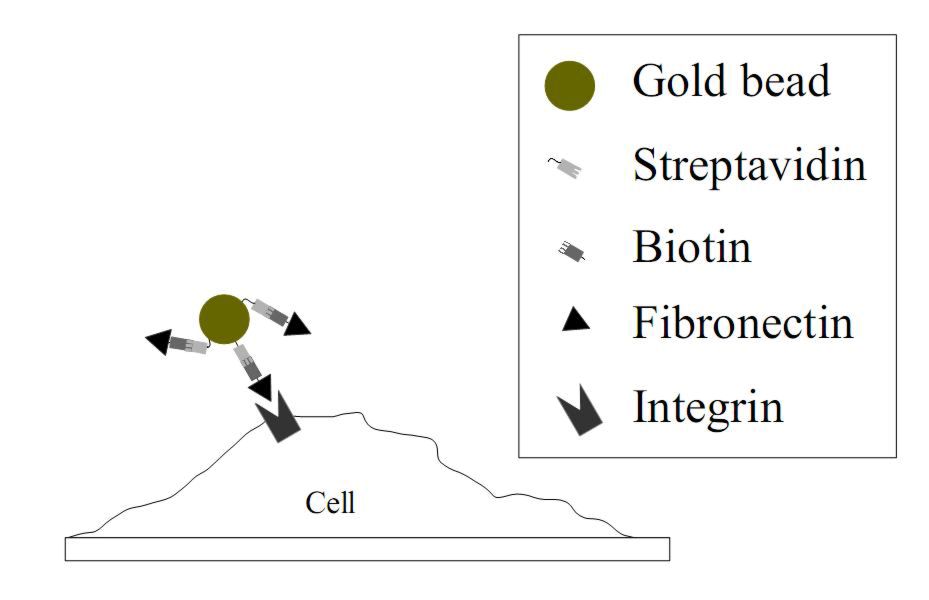}
\caption{Schematic representation  of the coupling between 3T3 cells
and gold beads}\label{function}
\end{figure}

We use NIH 3T3 mouse fibroblasts (quoted as 3T3 cells in this paper) with integrin surface receptors labeled with 40 nm
gold particles. Streptavidin-coated gold nanoparticles were attached to the cell surface integrin receptors via biotin
and fibronectin proteins: see Fig.~\ref{function}. Streptavidin and biotin are very well known for their strong
affinity towards each other, and fibronectin, an extracellular matrix protein, has the property of interacting
specifically with cell surface receptors of integrin family.

Fibronectin  proteins  (fibronectin from bovine plasma, Sigma, St Louis, MO) were labeled with biotin by using
EZ-Link$^{\textcircled {\scriptsize R}}$Sulfo-NHS-LC-Biotin according to the provider protocol (Pierce, Rockford, IL).
The final concentration of biotinylated-fibronectin solution was 0.447 mg/mL. We used a solution of gold nanoparticles
of 40 nm average diameter pre-coated with streptavidin. The streptavidin/gold conjugates (BioAssay, Gentaur, France)
were rinsed twice with 1 M PBS (Phosphate Buffered Saline) (pH~=~7.25) and 10 $\mu$L of the gold solution was added to
990 $\mu$L of the same PBS buffer to form 1 L dilute solution. Then the dilute solution was incubated with 50 $\mu$L
biotinylated-fibronectin solution for four hours at room temperature to allow streptavidin-biotin bonding. The
functionalized beads solution was stored at 4$^\circ$C and used within 24 hours after preparation in order to ensure
maximum functionality. Before every use, the functionalized beads solution was sonicated.

3T3 cells were cultured   in Duelbecco's modified Eagle's medium (DMEM Gibco, Invitrogen, Carlsbad, CA) supplemented
with 10\% fetal calf serum (PAA Laboratories GmbH)  on 32 mm diameter glass cover slips at 37$^\circ$C in a 5\% CO2
atmosphere, 48 hours before the observation. The cover slips were coated with fibronectin (fibronectin from bovine
plasma, Sigma, St Louis, MO) in advance for optimum cell growth. After being incubated for 24 hours, a monolayer of 3T3
fibroblasts was immersed into a solution composed of DMEM (2 mL) plus 500 $\mu$L of the functionalized beads solution.
The integrin-fibronectin link is created at this level allowing the cells to attach, on their surface, the
functionalised gold nanoparticles.

The coverslip  containing adherent 3T3 cells tagged with 40 nm gold nanoparticles was mounted on a specific observation
chamber. During the observation, cells were kept in DMEM F12 medium  with 0.5\% fetal calf serum with no phenolred, to
fulfill optimum survival conditions. The efficiency of biotinylation protocol is verified by measuring the level of
biotin incorporation  on an HABA [2-(4' -Hydroxyazobenzene)  Benzoic Acid)] quantitation assay. Average number obtained
of biotin molecules per fibronectin is 2.5.

\section{Experimental setup}\label{section_experimental_setup}

\begin{figure}
\centering\includegraphics[width=8cm]{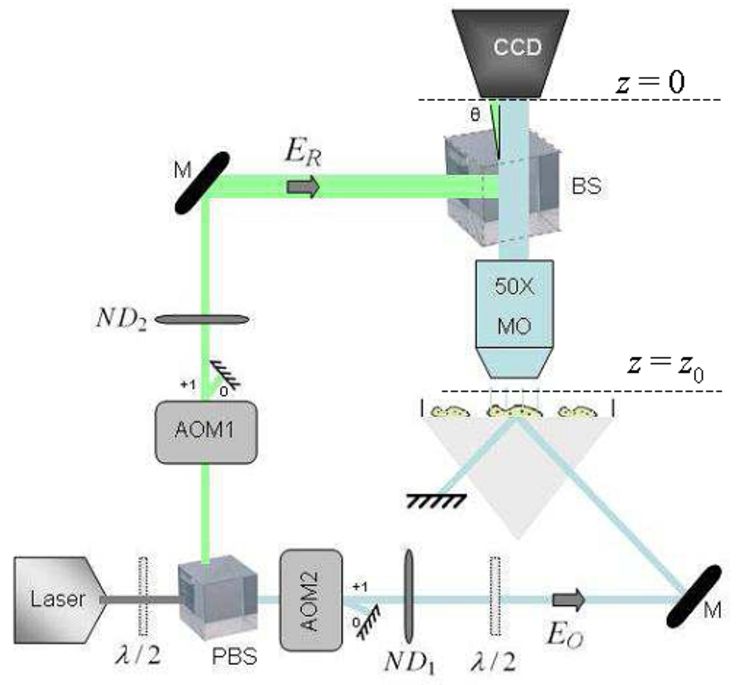} \caption{Experimental setup. AOM1, AOM2:  acousto-optic modulators;
M: mirror; MO: microscope objective NA = 0.5; $\lambda$/2: half wave plate; BS: beam splitter; PBS: polarizing beam splitter;
CCD: CCD camera; $\emph{E}_{R}$: reference wave; $\emph{E}_{O}$: object illumination wave; \emph{E}: scattered wave;
$\theta$: angular tilt; ${\emph{ND}}_{1}$, ${\emph{ND}}_{2}$: neutral density filters; $z=0$: CCD plane;   $z=z_0$: CCD
conjugate plane with respect to MO.}\label{expsetup}
\end{figure}

A schematic representation of the experimental setup is shown in Fig.~\ref{expsetup}. The illumination source is a
single mode near infrared laser diode emitting at $\lambda$=785nm (50 mW, 90 mA current). A
polarizing beam splitter cube (PBS) is used to split the original  illumination laser light into two beams, a reference
beam (complex field $\emph{E}_{R}$, frequency $\emph{f}_{R}$) and an object illumination beam (complex field
$\emph{E}_{O}$, frequency $\emph{f}_{O}$) forming the two arms of a Mach-Zehnder interferometer. A combination of a
half wave plate ($\lambda$/2) and two neutral density filters (${\emph{ND}}_{1}$, ${\emph{ND}}_{2}$) is used to prevent
the saturation of the detector by controlling the optical power traveling in each arm. Two acousto-optic modulators
(AOM1, AOM2) driven around 80 MHz with a selection of the first order of diffraction, shift both frequencies at
respectively ${f}_{AOM1}$ and ${f}_{AOM2}$ .

The object beam illuminates the sample,  prepared as described in
section \ref{section_Sample preparation}, by provoking total
internal reflection (TIR) at the cell/bead-air interface. The
evanescent wave locally frustrated by the beads and cells gives off
a propagating scattered wave (complex field \emph{E}), which is
collected by a microscopic objective (MO, $50 \times$ magnification,
NA=0.5, air). A beam splitter is then used to combine the scattered
object wave and the reference wave which is slightly angularly
tilted ($\theta\sim1^{\circ}$) with respect to the propagation axis
of the object wave in an off-axis configuration. The half wave plate
($\lambda$/2) on the object illumination arm aligns the polarization
of the corresponding beam allowing an optimal interference of the
beams. A CCD camera (Roper Cascade 512F, 512$\times$512 square
pixels  of 16$\mu$m size, exposure time 100 ms, frame rate
$f_{CCD}=8$ Hz) detects the interference pattern (hologram) and
sends it to a computer. The hologram is then numerically treated and
the complex field \emph{E(x,y,z)} is reconstructed numerically.

\section{Holographic  acquisition and   reconstruction }

A four-phase demodulation method is used  to record holograms. This method consists in acquiring a sequence of images
with a relative phase shift $\Delta\varphi=\pi/2$ between two consecutive frames. To get an accurate phase shift
\cite{atlan2007aps}, the reference wave is frequency shifted  by tuning the two acousto-optic modulators AOM1 and AOM2
\cite{Leclerc2000}, so that the heterodyne beat frequency is:
\begin{equation}\label{Eq_1a}
    \Delta f={f}_{AOM1}-{f}_{AOM2}=\frac{f_{CCD}}{4}
\end{equation}
where $f_{CCD}=$8Hz is the frame rate frequency of the CCD
camera. The camera records a sequence of 32 frames
$\emph{I}_{0}$,...,$\emph{I}_{31}$, and the object field $\emph{E}$
on the CCD plane ($z=0$) is given by:
\begin{equation}\label{eq2}
    \emph{E}(x,y,z=0) = \sum_{n=1}^{M}  j^{~n} \emph{I}_{n}
 \end{equation}
where $j^2=-1$, and $M$ is the number of frames used for the reconstruction, which is equal either to 1 or to 32 in the
experiment we present here.  In Eq.\ref{eq2}, the coordinates $x,y$ (with $0<x,y<511$) are integers, which represent
the pixel location within the CCD plane. The pixel size is then the physical CCD pixel size, i.e., 16 $\mu$m.

Since we image the sample through a microscope objective (MO), the measured hologram represents the
field $\emph{E}(x,y,z=z_0)$ within the CCD conjugate plane $z=z_0$ with respect to MO, i.e.,
the plane, whose image is on focus on the CCD detector. In that case, one must compensate the phase curvature, the
phase tilt and the enlargement factor that are related to the presence of MO \cite{colomb2006}. We have thus:
\begin{equation}\label{EQ_2b}
     \emph{E}(x,y,z=z_0)\;=\; e^{j (K_x x + K_y y)} \; e^{j A (x^2 + y^2) }  \sum_{n=0}^{M}  j^{~n} \emph{I}_{n}
\end{equation}
where $(K_x,K_y)$ and $A$ are the  tilt and lens parameters respectively. These parameters  are determined by
reconstructing the image of the microscope objective output pupil using the one Fourier transform method
\cite{schnars1994drh}.  The lens parameter is close   to the lens parameter that is used in the pupil reconstruction,
and the tilt parameter to the translation that pushes the pupil in the centre of the reconstructed image. The
magnification factor of the conjugate plane  is measured by imaging a USAF target. We get, in the conjugate plane,
pixel sizes of $\delta x,\delta y=177$ nm.

The object field $\emph{E}(x,y,z)$ is calculated  in the vicinity of conjugate plane (i.e. for $z \simeq z_0$) by the
angular spectrum method, which requires two Fourier transforms \cite{Leclerc2000,leclerc2001,yu2005}. This method is
chosen in order to keep the pixel size constant in the reconstruction. The reconstruction is done for 512 different
reconstruction distances
%
 $   z=z_0+ m \delta z $
%
where $\delta z=177$ nm and  $m=-256...+255 $. By  this way, we get
3D volume  images  with $512 \times 512\times 512$ voxels, with the
same pixel size (177 nm) in the 3 directions $x,y$ and $z$.

\section{Experimental Results}


\begin{figure}
    \centering\includegraphics[width=6cm]{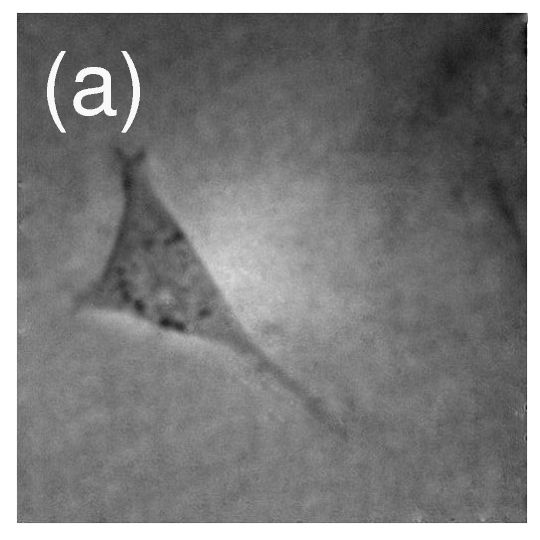}
   \centering\includegraphics[width=6cm]{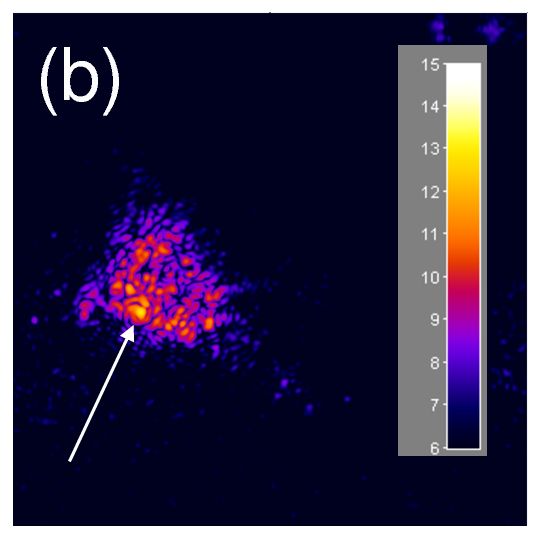}
  \centering\includegraphics[width=8cm]{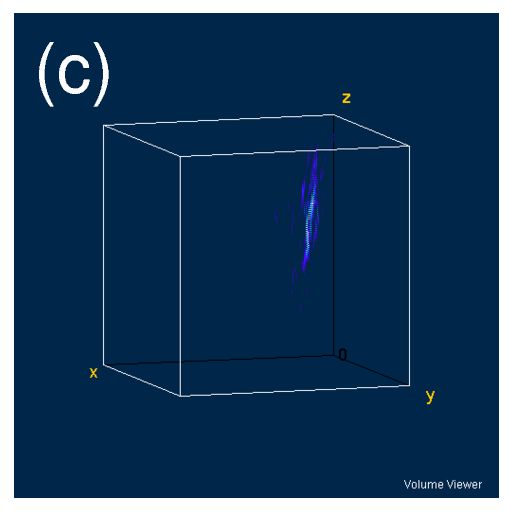}
 \caption{ Fibroblast cell tagged with a 40
nm gold particle. (a) Direct image under white light illumination. (b) Reconstructed Holographic intensity $I$ image.
The 40 nm gold particle is marked with a white arrow. The color scale corresponds to $6<\ln(I)<15$). (c) Volume view of
the 3D reconstructed data ($512 \times 512 \times 512$ voxels; voxel  size 177 nm in all directions). The holographic
reconstruction is made in (b) and (c) from 1 CCD frame with an exposure time of 100 ms. }\label{fig3new}
\end{figure}

We have considered a 3T3 cell with integrin cellular surface receptor tagged with a single 40 nm gold particle. Since we
are interested in tracking the particle, the camera exposure time is an important issue, and we will first reconstruct the 3D image of the sample using one CCD frame. Then, to improve the Signal to Noise Ratio and to confirm our results, we will
reconstruct the 3D image of the sample from the whole sequence of 32 frames. Figure \ref{fig3new} (a) shows a direct image of the sample under white
light illumination. While the cell is well defined on the left hand side of the image, the attached gold
particle cannot be seen.

Figure \ref{fig3new} (b) shows the reconstructed intensity image of the sample in a colored logarithmic scale. The
reconstruction is done using 1 CCD frame ($M=1$) with an exposure time of 100 ms. The reconstruction plane  is
$z=z_0+12.2~ \mu$m i.e. $n_z=325$ where $n_z$ is the z index on the 3D matrix, which is the plane where the brightest
point is detected. It also corresponds to the cell plane. The background
signal is the red and blue triangular
structure, with a few speckle yellow high spots. It is due to the light scattered by the inhomogeneities of the cell refractive index, and thus corresponds to the cell imaged in Fig.\ref{fig3new} (a). The gold particle corresponds to the brightest spot in Fig.\ref{fig3new} (b). It is located on
the left bottom of the cell, and marked with a white arrow.

To illustrate the ability of our technique to localize the gold particle in 3D, we have displayed in Fig. \ref{fig3new}
(c) a volume view  of the $512 \times 512 \times 512$  reconstructed intensity image 3D data, i.e., we have plotted
$I(x,y,z)=|E(x,y,z)|^2$ for different reconstruction $z$ distances. The bright light bluish zone on the 3D image
corresponds to the 40 nm gold particle signal. This wavefield has the shape of a
cigar roughly oriented along the microscope objective axis ($z$ axis).

\begin{figure}
     \centering\includegraphics[width=6cm]{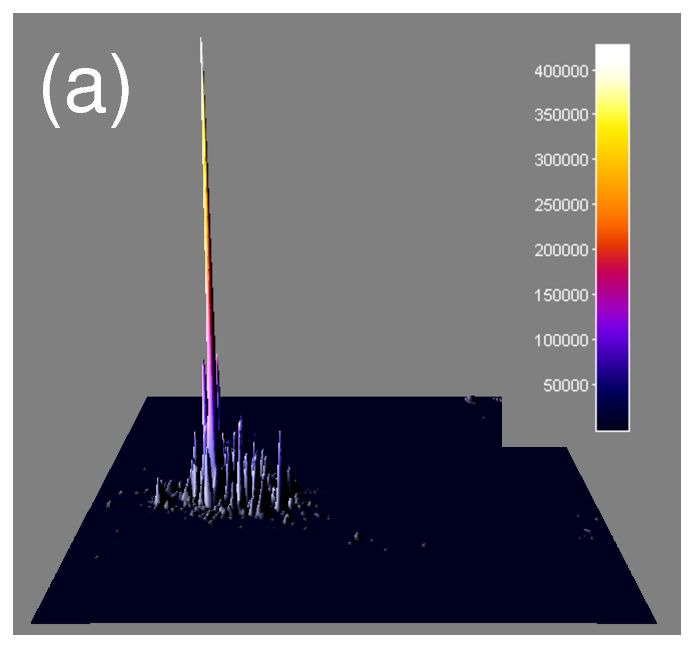}
    \includegraphics[width=7cm]{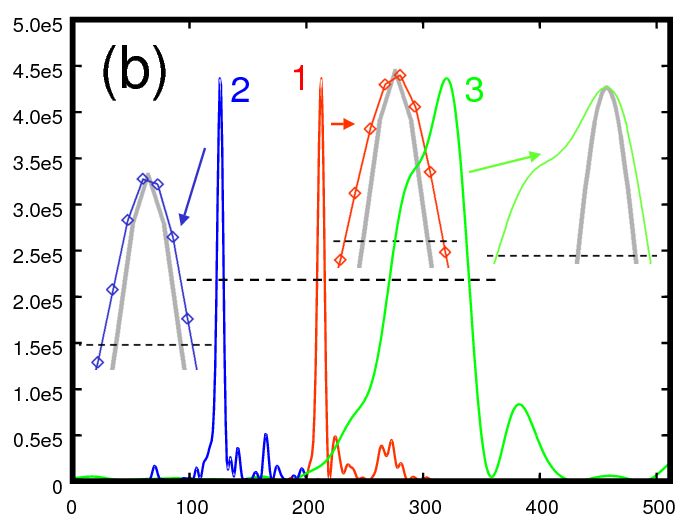}
 \caption{ A fibroblast cell tagged with a 40
nm gold particle using 1 frame acquisition. (a) 3D linear surface plot of the intensity image reconstructed from a
single frame. (b) Experimental linear plot of cuts made within the intensity signal $I$ along the three axis $x,y$ and
$z$ at the brightest voxel of Fig.\ref{fig3new} (b) corresponding to the gold nanoparticle location. Curves 1, 2 and 3
corresponds to the $x,y$ and $z$ axis respectively. The horizontal black dashed line represents the half maximum of the
curves. Curves marked with arrows are zooms of the corresponding curves from maxima to half maxima. Light grey curves
correspond to the different ideal theoretical curves. }\label{fig4new}
\end{figure}

To better visualize  the 40 nm gold particle, we have displayed, on Fig. \ref{fig4new} (a), a 3D linear surface plot of
the corresponding logarithmic scale colored image of Fig.\ref{fig3new} (b). As seen, the gold particle signal is much
higher than the background signal that corresponds to the speckle scattered by the cell. The particle can thus be
easily located.

To perform a quantitative analysis of the precision we expect in the 3D localization of the gold particle, we have
considered the 3D volume intensity image of Fig.\ref{fig3new} (c), and we have made cuts along the 3 axis ($x$, $y$ and
$z$) at the gold particle crossing point (i.e. where the three axis cross the highest intensity voxel of the 3D image).
The intensity signals $I$ along the 3 axis are plotted on Fig.\ref{fig4new} (b), curves 1 (red), 2 (blue) and 3 (green)
correspond to x, y and z axis respectively. The horizontal black dashed line indicates the half maximum of the curves. In order to
measure the curves widths, we have displayed zooms of the curves from maximum to half maximum. The individual pixels
are visible on curves 1 and 2 zooms (x and y axis). The Full Width at Half Maximum (FWHM) is about 6 pixels in the x
and y directions ($6 \times 177 = 1060$ nm), and about 60 pixels in the z direction (10.6 $\mu$m).

To compare the resolution   obtained here with that expected from the NA of the microscope objective, we have
computed the wavefield $E_I(x,y,z)$ that is expected in the ideal case for a NA=0.5 objective, and we have made cuts
along x, y and z directions of the $512\times 512\times 512$ ideal cube of data for  $|E_I(x,y,z)|^2$. The
corresponding theoretical cuts are plotted with the zooms in heavy grey line. Like for the experimental cuts, the
horizontal axis is in pixels. In the x and y directions, the obtained FWHM is close to the expected one. Yet, in
the z direction the shape of the curve is not symmetric, and the measured FWHM is about 3 times larger than the
expected FWHM in the ideal case. This discrepancy may be related to the parasitic speckle background signal.

Using a parabolic approximation for the local field, the location of the gold particle can be calculated  by fitting
the data points that are above half maximum. 
The accuracy of the measurement made by this method is $\pm 5$ nm in the x and y directions (uncertainty given by the fit software Gnuplot). It corresponds also to the variation of the particle when the number of data points used for the fit is increased from 8 to 7. In the z direction, the
measured curve is not perfectly symmetric, and so the fit technique is not very accurate because the result of the fit strongly depends on the data points that are used in the calculation.

\begin{figure}
     \centering\includegraphics[width=6cm]{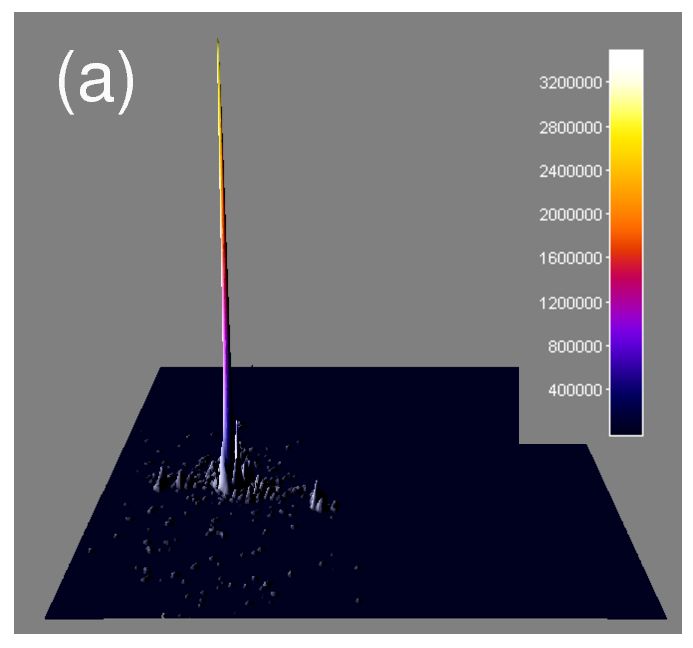}
   \includegraphics[width=6cm]{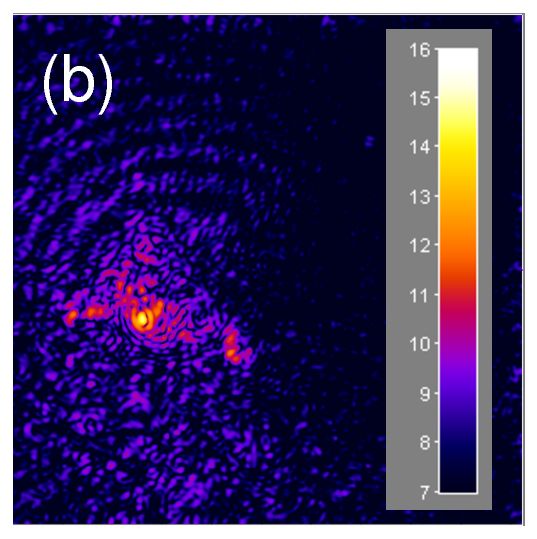}
 \caption{ A fibroblast cell tagged with a 40
nm gold particle using 32 frames acquisition. (a) 3D linear surface plot of the 32 frames reconstructed holographic
intensity image. (b) Reconstructed Holographic intensity $I$ image. The 40 nm gold particle is marked with a white
arrow. The color scale corresponds to $6<\ln(I)<15$). }\label{fig5new}
\end{figure}

In order to visualize the gold particle with a better Signal to Noise Ratio (SNR), we have also reconstructed the
holographic images using a sequence of $M=32$ acquired CCD frames. In this case, the total exposure time is $32
\times 100$~ms=3.2 s, while the measurement time is $32 / f_{CCD}$ = 4 s. Figure \ref{fig5new} (a) shows the 3D linear
surface plot of the reconstructed intensity image in the $z=z_0+6.1 \mu$m plane. Figure \ref{fig5new} (a) is similar to Fig.\ref{fig3new} (a) but obtained with 32 frames, instead of 1 frame. Here again, not only the gold particle signal is much
higher than the background signal, but the ratio of the particle signal versus background signal (light scattered by the cell) is increased with
respect to the ratio obtained for Fig.\ref{fig4new} (a), and the visibility of the gold particle is improved. This
visibility improvement is confirmed by Fig. \ref{fig5new} (b), which shows the reconstructed intensity image of the
sample in a colored logarithmic scale. The cell signal, corresponding to the triangular bluish and reddish structure, is
still visible. The particle signal in yellow is visible too, but with a much better contrast than in Fig.\ref{fig4new} (a).

\begin{figure}[h]
    \centering\includegraphics[width=6.3cm]{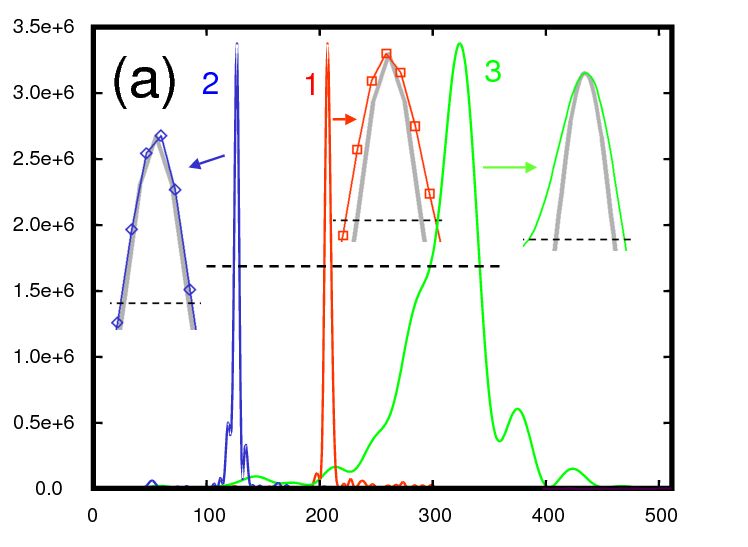}
   \centering\includegraphics[width=6cm]{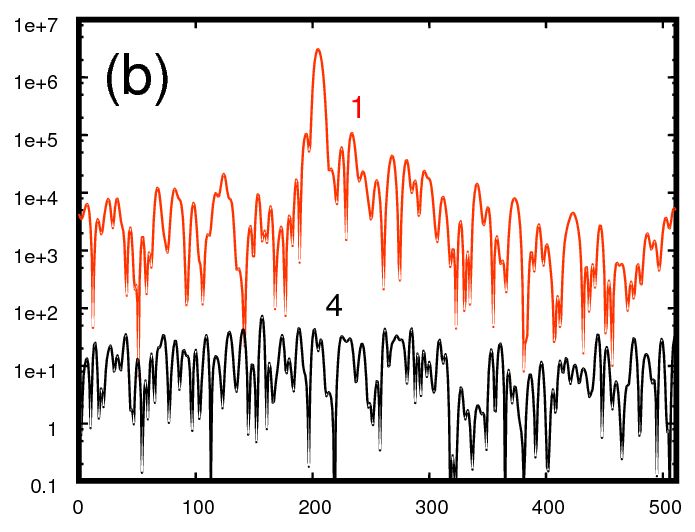}
\caption{ (a) Experimental linear plot  of cuts made within the intensity signal $I$ along the three axis $x,y$ and $z$
at the brightest voxel of Fig.\ref{fig5new} (b) corresponding to the gold nanoparticle location. Curves 1, 2 and 3
corresponds to the $x,y$ and $z$ axis respectively. The horizontal black dashed line represents the half maximum of the
curves. Curves marked with arrows are zooms of the corresponding curves from maximum to half maximum. Light grey curves
correspond to the different ideal theoretical curves. (b) Intensity plot of the cut along the x axis in logarithmic
scale 1 at the nanoparticle location and 4 without illuminating the sample }\label{fig6new}
\end{figure}

Figure \ref{fig6new} (a) shows the cuts made along the 3 axes ($x$, $y$ and $z$) obtained for the 32 frames data. Here
again,  we have displayed zooms of the three cross sections from maxima to half maxima as well as zooms of the
theoretical curves. These cross sections are similar to the cross sections obtained in the case of 1 frame
acquisition, but in this case (32 frames) the heavy grey lines corresponding to the theoretical zoomed curves are
closer to the experimental curves. This confirms our hypothesis of the effect of the background parasitic
signal (which is much lower here, as seen by comparing Fig.\ref{fig5new} (a) with Fig.\ref{fig4new} (a)). Here again, the location of the particle can be calculated by fitting the measured data point with a parabola. The
accuracy is still $\pm 5$ nm is the x and y directions, but since the curve 3 is now roughly symmetric, the fit can also be performed in the z
direction, and the particule can be located with an accuracy estimated to $\pm 100$ nm in the z direction.

To better compare  the particle signal with respect to the light scattered by the cell (background speckle), we plotted
the intensity signal along x (which corresponds to curve 1 of Fig.\ref{fig6new} (a) ) in logarithmic scale on
Fig.\ref{fig6new} (b). As seen, the particle signal is 2 to 3 orders of magnitude larger than the background signal. It
has the same order of magnitude that the signal obtained in a control experiment of 40 nm gold particles within a water and agarose suspension. We plotted also (black curve 4) the background signal obtained without illuminating the
sample. This background corresponds to the ultimate sensitivity limit of the holographic detection
\cite{gross2007digital}, i.e. to the shot noise on the local oscillator beam. As seen, the shot noise background  is
about $100 \times$ lower than the cell background. This means that if the exposure time is reduced by the same factor
(from 3.2 s down to 32 ms), one expects to get roughly the same Signal to Noise Ratio (SNR$\simeq 100$) on the particle
signal. Since the illumination area (about 1 mm$^2$) is much larger than the imaged region, it is possible, by
focusing the illumination, to reduce further the exposure time without significally decreasing the SNR.

\section{Conclusion}

In conclusion, we have reported in this article a digital holographic microscopy technique capable of imaging 40 nm
gold nanoparticles attached to living cells surface receptors. Results show that this method can be effectively used
to distinguish nanoparticles from the cellular structures. We show in particular that the 40 nm gold particle
scattered signal is much larger in intensity than the field scattered by the  cell structures. While a single acquired
image is sufficient to distinguish the signal of the particle from the cell scattered field, better signal to noise
ratio and z-positions estimations can be obtained if the reconstruction is done by averaging on a sequence of several acquired images. Overall, we showed that we can use our technique for the 3D mapping of the sample's full scattered field using a single recorded
hologram and without any mechanical scanning.

\acknowledgments     

he authors wish to acknowledge the French National Research Agency (ANR)and the ``Centre de Comp\'{e}tence NanoSciences
\^{I}le de France''(C'nano IdF) for their support.



\begin{thebibliography}{99}
\newcommand{\enquote}[1]{``#1''}
\expandafter\ifx\csname url\endcsname\relax
  \def\url#1{\texttt{#1}}\fi
\expandafter\ifx\csname urlprefix\endcsname\relax\def\urlprefix{URL }\fi
\providecommand{\eprint}[2][]{\url{#2}}

\bibitem{jain2006}
P.~K. Jain, K.~S. Lee, I.~H. El-Sayed, and M.~A. El-Sayed, \enquote{{Calculated
  absorption and scattering properties of gold nanoparticles of different size,
  shape, and composition: Applications in biological imaging and biomedicine},}
  J. Phys. Chem. B \textbf{110}, 7238--7248 (2006).

\bibitem{west06}
J.~L. West, R.~A. Drezek, and H.~N. J., \enquote{Nanotechnology provides new
  tools for biomedical optics,} in \emph{Tissue Engineering and Artifical
  Organs, 3rd Edition}, J.~D. Bronzino, ed., pp. 25--1--25--9 (CRC Press,
  2006).

\bibitem{lasne2006}
D.~Lasne, G.~A. Blab, S.~Berciaud, M.~Heine, L.~Groc, D.~Choquet, L.~Cognet,
  and B.~Lounis, \enquote{{Single nanoparticle photothermal tracking (SNaPT) of
  5-nm gold beads in live cells},} Biophys. J. \textbf{91}, 4598--4604 (2006).

\bibitem{cognet2002}
L.~Cognet, C.~C.~Tardin, D.~Boyer, D.~Choquet, P.~Tamarat, and B.~Lounis,
  \enquote{{Single metallic nanoparticles imaging for protein detection in
  cells},} in \emph{Proc. Natl. Acad. Sci.} (USA).

\bibitem{boyer2003}
D.~Boyer, P.~Tamarat, A.~Maali, B.~Lounis, and M.~Orrit, \enquote{{Photothermal
  imaging of nanometer-sized metal particles among scatterers},} Science
  \textbf{297}, 1160--1163 (2003).

\bibitem{raschke2003}
G.~Raschke, S.~Kowarik, T.~Franzel, C.~Sonnichsen, T.~A. Klar, and J.~Feldmann,
  \enquote{{Biomolecular recognition based on single gold nanoparticles light
  scattering},} Nano Lett. \textbf{3}, 935--938 (2003).

\bibitem{elsayed2005spr}
I.~H. El-Sayed, X.~Huang, and M.~A. El-Sayed, \enquote{{Surface plasmon
  resonance scattering and absorption of anti-EGFR antibody conjugated gold
  nanoparticles in cancer diagnostics: Applications in oral cancer},} Nano
  Lett. \textbf{5}, 829--834 (2005).

\bibitem{Schnars_2002}
U.~Schnars and W.~P.~O. J{\"u}ptner, \enquote{Digital recording and numerical
  reconstruction of holograms,} Meas. Sci. Technol. \textbf{13}, R85--R101
  (2002).

\bibitem{atlan2007aps}
M.~Atlan, M.~Gross, and E.~Absil, \enquote{{Accurate phase-shifting digital
  interferometry},} Opt. Lett. \textbf{32}, 1456--1458 (2007).

\bibitem{leclerc2001}
F.~LeClerc, M.~Gross, and L.~Collot, \enquote{{Synthetic-aperture experiment in
  the visible with on-axis digital heterodyne holography},} Opt. Lett.
  \textbf{26}, 1550--1552 (2001).

\bibitem{di2008}
J.~Di, J.~Zhao, H.~Jiang, P.~Zhang, Q.~Fan, and W.~Sun, \enquote{{High
  resolution digital holographic microscopy with a wide field of view based on
  a synthetic aperture technique and use of linear CCD scanning},} Opt. Lett.
  \textbf{47}, 5654--5659 (2008).

\bibitem{carl2004}
D.~Carl, B.~Kemper, G.~Wernicke, and G.~von Bally,
  \enquote{{Parameter-Optimized Digital Holographic Microscope for
  High-Resolution Living-Cell Analysis},} Appl. Opt. \textbf{43}, 6536--6544
  (2004).

\bibitem{gross2007dhu}
M.~Atlan, M.~Gross, and E.~Absil, \enquote{{Digital holography with ultimate
  sensitivity},} Opt. Lett. \textbf{32}, 909--911 (2007).

\bibitem{Leclerc2000}
F.~LeClerc, L.~Collot, and M.~Gross, \enquote{Numerical heterodyne holography
  with two-dimensional photo-detector arrays,} Opt. Lett. \textbf{25}, 716--718
  (2000).

\bibitem{cuche2000sfz}
E.~Cuche, P.~Marquet, C.~Depeursinge, \emph{et~al.}, \enquote{{Spatial
  filtering for zero-order and twin-image elimination in digital off-axis
  holography},} Applied Optics \textbf{39}(23), 4070--4075 (2000).

\bibitem{Leith65}
E.~Leith and J.~Upatnieks, \enquote{Microscopy by wave front reconstruction,}
  J. Opt. Soc. Am. \textbf{55}, 981--986 (1965).

\bibitem{yamaguchi1997psd}
I.~Yamaguchi and T.~Zhang, \enquote{{Phase-shifting digital holography},} Opt.
  Lett. \textbf{22}, 1268--1270 (1997).

\bibitem{gross2008naa}
M.~Gross, M.~Atlan, and E.~Absil, \enquote{{Noise and aliases in off-axis and
  phase-shifting holography},} Appl. Opt. \textbf{47}, 1757--1766 (2008).

\bibitem{xu2001}
W.~Xu, M.~H. Jericho, I.~A. Melnertzhagen, and H.~J. Kreuzer, \enquote{{Digital
  in-line holography for biological applications},} in \emph{Proc. Natl. Acad.
  Sci.} (USA).

\bibitem{mann2005}
C.~J. Mann, L.~Yu, C.~M. Lo, and M.~K. Kim, \enquote{{High resolution
  quantitative phase-contrast microscopy by digital holography},} Opt. Express
  \textbf{13}, 8693--8698 (2005).

\bibitem{mann2006}
C.~J. Mann, L.~Yu, and M.~K. Kim, \enquote{{Movies of cellular and sub-cellular
  motion by digital holographic microscopy},} Biomed. Eng. Online \textbf{5},
  21 (2006).

\bibitem{charrire2006}
F.~Charri{\'e}re, A.~Marian, F.~Montfort, J.~Kuehn, and T.~Colomb,
  \enquote{{Cell refractive index tomography by digital holographic
  microscopy},} Opt. Lett. \textbf{31}, 178--180 (2006).

\bibitem{atlan2008}
M.~Atlan, M.~Gross, P.~Desbiolles, E.~Absil, G.~Tessier, and M.~Coppey-Moisan,
  \enquote{{Heterodyne holographic microscopy of gold particles},} Opt. Lett
  \textbf{35}, 500--502 (2008).

\bibitem{absil2009photothermal}
E.~Absil, G.~Tessier, M.~Gross, M.~Atlan, N.~Warnasooriya, S.~Suck,
  M.~Coppey-Moisan, and D.~Fournier, \enquote{{Photothermal heterodyne
  holography of gold nanoparticles},} Opt. Express (to be published)  (2009).

\bibitem{goldberg1986saf}
D.~Goldberg and D.~Burmeister, \enquote{{Stages in axon formation: observations
  of growth of Aplysia axons in culture using video-enhanced
  contrast-differential interference contrast microscopy},} Journal of Cell
  Biology \textbf{103}(5), 1921--1931 (1986).

\bibitem{colomb2006}
T.~Colomb, F.~Montfort, J.~K\"{u}hn, N.~Aspert, E.~Cuche, A.~Marian,
  F.~Charri{\'e}re, S.~Bourquin, P.~Marquet, and C.~Depeursinge,
  \enquote{{Numerical parametric lens for shifting, magnification, and complete
  aberration compensation in digital holographic microscopy},} J. Opt. Soc. Am.
  A \textbf{23}, 3177--3190 (2006).

\bibitem{schnars1994drh}
U.~Schnars and W.~J{\"u}ptner, \enquote{{Direct recording of holograms by a CCD
  target and numerical reconstruction},} Applied Optics \textbf{33}(2),
  179--181 (1994).

\bibitem{yu2005}
L.~Yu and M.~Kim, \enquote{{Wavelength-scanning digital interference holography
  for tomographic three-dimensional imaging by use of the angular spectrum
  method},} Opt. Lett. \textbf{30}, 2092--2094 (2005).

\bibitem{gross2007digital}
M.~Gross and M.~Atlan, \enquote{{Digital holography with ultimate
  sensitivity},} Optics Letters \textbf{32}(8), 909--911 (2007).

\end{thebibliography}

\end{document}